\documentclass{jfm}
\usepackage{natbib}
\usepackage{epsfig}
\usepackage{amssymb}
\def\begineq{\begin{equation}}
\def\endeq{\end{equation}}
\def\be{\begin{equation}}
\def\ee{\end{equation}}

\title[Homogeneous shear turbulence]{Derivative moments in
stationary homogeneous shear turbulence}
\author[J. Schumacher]%
{J\ls \"O\ls R\ls G\ns S\ls C\ls H\ls U\ls M\ls A\ls C\ls H\ls E\ls R}
\affiliation{Department of Mechanical Engineering,
Yale University,
New Haven, CT 06520-8284, USA\\[\affilskip]
Fachbereich Physik, Philipps-Universit\"at,
D-35032 Marburg, Germany}
\pubyear{2001}
\volume{999}
\pagerange{1--10}
\date{?? and in revised form ??}
\begin{document}
\maketitle
\begin{abstract}
A statistically stationary and nearly homogeneous turbulent shear
flow is established by an additional volume forcing in combination
with stress-free boundary conditions in the shear direction. Both
turbulent energy and enstrophy are stationary to a much better
approximation than in previous simulations that use remeshing. The
temporal fluctuations decrease with increasing Reynolds number. Energy
spectra and shear-stress cospectra show that local isotropy
is satisfactorily obeyed at the level of second-order moments.
However, derivative moments of high-order up to $n=7$ yield
increasing moments for $n\ge 4$ for the spanwise vorticity and the
transverse derivative of the streamwise velocity in the range of
Taylor Reynolds numbers $59\le R_{\lambda}\le 99$. These findings,
which are in apparent violation of local isotropy, agree with
recent measurements.
\end{abstract}

\section{Introduction}
Turbulent flows in nature and laboratory are mostly anisotropic on their largest
scales.  A typical situation is one in which an imposed shear gradient sustains
the turbulence.  One fundamental question is the effect of anisotropic large
scales on the statistical behaviour on the smaller scales.  Kolmogorov (1941)
postulated that at sufficiently large Reynolds number statistics of small scales
become isotropic even if the large scale driving is anisotropic.  Consequently,
cross-correlations, indicated, e.g., by the shear-stress cospectrum $E_{xy}(k)$
in the Fourier space, should decay very fast to zero with respect to increasing
wavenumber.  Based on this postulate, Lumley (1967) predicted, for a turbulent
shear flow with constant mean shear rate $S$, the rate at which anisotropy
vanishes with decreasing scale size.  If the shear time scale, $T_s=S^{-1}$, is
much greater than the viscous time scale, $\tau_{\eta}=(\nu/\epsilon)^{1/2}$, in
other words, if $S\tau_{\eta}\ll 1$ holds, he showed by dimensional arguments
that the shear-stress cospectrum follows $E_{xy}(k)\sim\epsilon^{1/3} S
k^{-7/3}$.  Quantities $\nu$ and $\epsilon$ are the kinematic viscosity and the
energy dissipation rate, respectively.  Note that the cospectrum would be
exactly zero in a system that is perfectly isotropic on all scales.  The concept
of local isotropy on the level of second order moments seemed to be supported by
direct numerical simulations (She {\it et al.}  1993) and a number of
high-Reynolds number measurements in turbulent boundary layers (Caughey {\it et
al.}  1979; Saddoughi \& Veeravalli 1994).  Very recently, however, Kurien \&
Sreenivasan (2000) found a slower decay of anisotropies in an atmospheric
boundary layer at $R_{\lambda}\approx 2000$.  They reported, e.g., a scaling of
the cospectrum with $E_{xy}(k_x)\sim k_x^{-2.1}$.  Deviations from the local
isotropy were also found in a number of new experiments in nearly homogeneous
shear flows when attention was focused on higher order moments (Garg \& Warhaft
1998; Ferchichi \& Tavoularis 2000; Shen \& Warhaft 2000).  

Quantities which are very sensitive to anisotropic contributions at smallest
scales are derivative moments of the turbulent velocity components, $u'_i$.  We
denote moments of odd orders ($k=1, 2, 3$) by
%-------------------------------------------------------------------------------
\begin{equation}
S_{2k+1}(\partial u'_i/\partial x_j)=
\frac{\langle (\partial u'_i/\partial x_j)^{2k+1}\rangle}
{\langle (\partial u'_i/\partial x_j)^{2}\rangle^{k+1/2}}\,,
\end{equation}
and of even orders ($k=2, 3$) by
\begin{equation}
K_{2k}(\partial u'_i/\partial x_j)=
\frac{\langle (\partial u'_i/\partial x_j)^{2k}\rangle}
{\langle (\partial u'_i/\partial x_j)^{2}\rangle^{k}}\,.
\end{equation}
%-------------------------------------------------------------------------------
The brackets $\langle\cdot\rangle$ denote the ensemble average and no tensor
summations are applied for $i$ and $j$.  If the turbulence in the shear flow is
locally isotropic odd moments of the transverse derivative ($i=x, j=y$) should
decay with $R_{\lambda}^{-1}$.  Shen \& Warhaft (2000) measured a slower decay
of the derivative skewness, $S_3(\partial u'_x/\partial y)\sim
R_{\lambda}^{-0.5}$, with a value of 0.2 at $R_{\lambda}\sim 1000$. The fifth
moment $S_5(\partial u'_x/\partial y)$ remained constant while $S_7(\partial
u'_x/\partial y)$ even increased in the range of Taylor Reynolds numbers between
100 and 1000.  These results were confirmed in part by similar experiments of
Ferchichi \& Tavoularis (2000).  Third order transverse derivative
moments of both experiments agreed also with those found in homogeneous shear
flow simulations at $R_{\lambda}\sim 100$ (Pumir 1996).  All the results
indicate that the influence of large scale anisotropies on the statistics of the
smallest scales does not decay as fast as predicted by dimensional arguments,
and in some cases may not decay at all.

This paper is addressed to the numerical investigation of statistical properties
in the simplest shear flow configuration, the homogeneous shear flow.  The flow
is characterized by the following mean profiles of the velocity components
%-------------------------------------------------------------------------------
\begin{equation} 
\langle u_x\rangle=S y,\,\,\,\langle u_y\rangle=\langle
u_z\rangle=0, 
\end{equation} 
%-------------------------------------------------------------------------------
where $x$ are streamwise (or downstream), $y$ shear (or wall-normal), and $z$
spanwise directions, respectively.  The velocity components can be decomposed in
a mean fraction and a fluctuating turbulent part, $u_i=\langle u_i\rangle+u'_i$
for $i=x, y$ and $z$.  Besides finite difference methods with shear periodic
boundary conditions (Gerz {\it et al.}  1989) pseudospectral methods in a fully
periodic domain using a remeshing (Rogallo 1981; Rogers \& Moin 1987; Pumir
1996; Gualtieri {\it et al.}  2000) were mostly used to model a homogeneous shear
flow.  In this method, it was harder to reach a stationary turbulent state.  An
alternative method to overcome this problem was suggested by Schumacher \&
Eckhardt (2000) very recently.  The method avoids the remeshing that needs to be
applied for the increasingly skewed grid of Rogallo (1981) and that
corresponds to a
time-periodic driving of the flow.  A statistically stationary state for long
times can be maintained by using stress-free boundary conditions in $y$
direction and an appropriate body force.  Results for third and fourth order
agreed with those of Rogers \& Moin (1987) and of Pumir (1996).  Derivative
skewness was weakly decaying with respect to $R_{\lambda}$.  This paper extends
that analysis.  We systematically investigate the temporal fluctuations of the
energy and enstrophy around their mean values and compare them with the findings
of Pumir (1996) and Gualtieri {\it et al.}  (2000).  We show that the
level of fluctuations around the mean quantities is much smaller and reaches 
values that are observed in turbulence measurements.  Energy as well as
shear-stress cospectra are studied.  Inspired by recent experiments we extend
the analysis systematically to higher order derivative moments ($n\le 7$) in the
range of Taylor Reynolds numbers accessible.  Non-decreasing higher order
moments indicate that very rare small-scale events may exist.  We shall also
discuss the problem of statistical convergence of higher order moments in
detail.

\section{Numerical model}
With length scales measured in units of the gap width $L_y$, and time scales
in units of $S^{-1}$, the dimensionless form of the
equations for an incompressible Navier--Stokes fluid become
%-------------------------------------------------------------------------------
\begin{equation}
\label{nseq}
\frac{\partial{\bf u}}{\partial t}+({\bf u}\cdot{\bf \nabla}){\bf u}
=-{\bf \nabla} p+\frac{1}{Re_s}{\bf \nabla}^2{\bf u}+{\bf f}\;,\;\;\;\;\;\;
{\bf \nabla}\cdot{\bf u}=0\;,
\end{equation}
%-------------------------------------------------------------------------------
where $p({\bf x},t)$ is the pressure, ${\bf u}({\bf x},t)$ the velocity field.
The shear Reynolds number is $Re_s=S\,L_y^2/\nu$.  In the $x$ and $z$ directions
periodic boundary conditions apply.  In the other direction the flow domain is
bounded by two parallel flat surfaces that are assumed to be impenetrable and
stress-free,
%-------------------------------------------------------------------------------
\begin{equation}
u_y=\frac{\partial u_x}{\partial y}=\frac{\partial u_z}{\partial y}=0
\quad\mbox{for}\quad y=0,\;L_y\,.
\end{equation}
%-------------------------------------------------------------------------------

As discussed in more detail in a recent paper by Schumacher \& Eckhardt (2001),
the free slip boundary conditions allow for efficient numerical simulations with
Fourier modes for the velocity components.  The equations are integrated by
means of a pseudospectral technique using a 2/3-rule de-aliasing.  Time
integration was done by means of a 5-th order Runge-Kutta scheme with adaptive
time stepping (Hairer {\it et al.}  1993).  In table 1 the computational
parameters are summarized for all the runs presented.  The aspect ratio
$L_x/L_z$ and the shear Reynolds number $Re_s$ were varied.  It is also visible
that the spectral resolutions in $x$ and $y$ directions are different causing
maximum wavenumber components of $k_{x,max}(=k_{z,max})=2\pi N_x/(3 L_x)$ and
$k_{y,max}=2\pi N_y/3$, respectively, after de-aliasing.  As a criterion for
sufficient spectral resolution $k_{max} \eta \ge 1$ is used 
(Pope 2000) with Kolmogorov length scale $\eta=(\nu^3/\epsilon)^{1/4}$ and
$k_{max}=\sqrt{8}\pi N_x/(3 L_x)$ for a uniform grid case.
For the present model with different grid spacing in the three dimensions, 
it is suggested therefore to take 
$k_{max}=((k_{x,max}^2+k_{y,max}^2+k_{z,max}^2)/3)^{1/2}$.
Additionally we plotted the ratio of grid spacing in shear direction to the
Kolmogorov scale, $\Delta y/\eta$, to indicate the sufficient spectral 
resolution.
%-----------------------------------------------------------------------
\begin{table}
\begin{center}
\begin{tabular}{lccccc}
Run No. & 1 & 2 & 3 & 4 & 5 \\
\hline
$Re_s$ & 500 & 1000 & 1000 & 2000 & 2000 \\
$R_{\lambda}$ & 59 & 79 & 87 & 95 & 99 \\
$L_x/L_y$ & $2\pi$ & $4\pi$ & $2\pi$ & $4\pi$ & $2\pi$ \\
$L_z/L_y$ & $2\pi$ & $2\pi$ & $2\pi$ & $2\pi$ & $2\pi$ \\
$N_x$ & 256 & 256 & 256 & 256 & 256 \\
$N_y$ & 65 & 129 & 65 & 129 & 65 \\
$N_z$ & 256 & 128 & 256 & 128 & 256 \\
$k_{max} \eta$ & 2.82 & 2.48 & 1.62 & 1.54 & 1.00 \\
$\Delta y/\eta$ & 0.57 & 0.50 & 0.99 & 0.80 & 1.60 \\
$\lambda/\eta$ & 17.8 & 18.9 & 19.8 & 20.2 & 20.7 \\
$L/\eta$ & 36.4 & 63.3 & 58.4 & 80.5 & 77.5 \\
\end{tabular}
\label{tab1}
\caption{List of the presented runs with its parameters: shear Reynolds
number $Re_s$, Taylor
Reynolds number $R_{\lambda}$, aspect ratios $L_x/L_y$ and $L_z/L_y$, the
numbers of grid points
$N_x$, $N_y$, and $N_z$, the spectral resolution criterion $k_{max} \eta$,
and the ratios of grid
spacing in $y$ direction $\Delta y$, of Taylor microscale $\lambda$, and of integral
length scale $L$ to
Kolmogorov length $\eta$.}
\end{center}
\end{table}
%---------------------------------------------------------------------------
The Taylor microscale $\lambda$ is defined here with the
downstream root-mean-square velocity
%-------------------------------------------------------------------------------
\begin{equation}
\lambda=\frac{\langle(u'_x)^{2}\rangle^{1/2}}
{\langle(\partial u'_x/\partial x)^2\rangle^{1/2}}
=\frac{u'_{x, rms}}
{(\partial u'_x/\partial x)_{rms}}\,,
\end{equation}
%-------------------------------------------------------------------------------
and the integral length scale $L$ is given by
%-------------------------------------------------------------------------------
\begin{equation}
L=\frac{1}{\langle(u'_x)^2\rangle}\int_0^{\infty} dr
\langle u'_x({\bf x}+r{\bf e_x})\,u'_x({\bf x})\rangle\,.
\end{equation}
%-------------------------------------------------------------------------------
The Taylor Reynolds number follows as $R_{\lambda}=u'_{x, rms}\lambda/\nu$.
%---------------------------------------------------------------------------
\begin{figure}
\vspace{0.5cm}
\epsfxsize=0.5\hsize
{\hskip 0.0cm{\centerline{\epsfbox{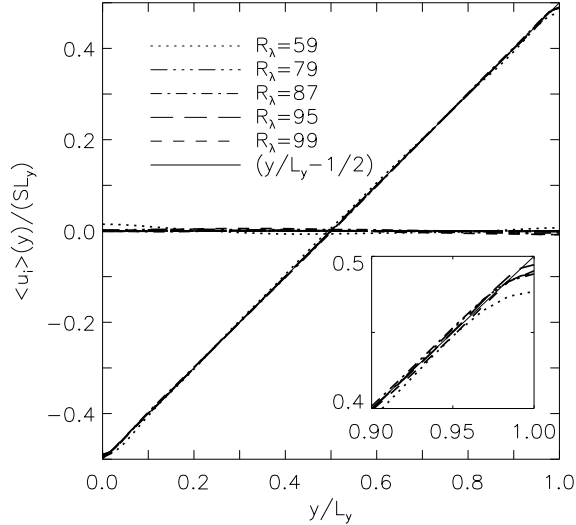}}}}
\vspace{0.5cm}
\caption{Mean velocity profiles of the three velocity components for
different Reynolds numbers. The inset magnifies the profiles at the
boundary $y/L_y=1$ where deviations from transverse homogeneity are present.
The thin solid line corresponds with the exact linear profile.}
\label{fig1}
\end{figure}
%-----------------------------------------------------------------------
The mean shear and turbulence are maintained by a suitable body force ${\bf
f}({\bf x},t)$. An almost linear mean profile $\langle u_x\rangle(y)=(y-1/2)$
for $y\in [0,1]$ (in dimensionless form) can be approximated by a finite
Fourier
sum of cosines
%-------------------------------------------------------------------------------
\begin{eqnarray}
\langle u_x\rangle(y)\simeq - \frac{4}{\pi^2}\sum^{5}_{n=0}
\frac{\cos[(2n+1)\pi y]}{(2n+1)^2}\,.
\label{profile}
\end{eqnarray}
%-------------------------------------------------------------------------------
The external forcing ${\bf f}$ was chosen such that the six modes used in
(\ref{profile}) remained constant in time, {\it i.e.}  $\partial \mbox{Re}
\{u_x({\bf q},t)\}/\partial t=0$ for Fourier modes with ${\bf q}=(2n+1)\pi {\bf
e}_y$ for $n=0$ to 5.  It was shown by Schumacher \& Eckhardt (2000) that this
forcing, although itself varying with respect to time, causes the expected mean
profiles of the velocity field components, $\langle u_y\rangle = \langle
u_z\rangle=0$ and $\langle u_x\rangle=y$ (see figure~\ref{fig1}).  The effects
of the free slip surfaces at $y=0$ and $y=L_y$ on the bulk behaviour are much
weaker than those of rigid walls, since only the wall-normal component is forced
to vanish and a nearly homogeneous, statistically stationary turbulent state can
be established.

\section{Energy balance and power spectra}
\begin{figure}
\vspace{0.5cm}
\epsfxsize=0.9\hsize
{\hskip 0.0cm{\centerline{\epsfbox{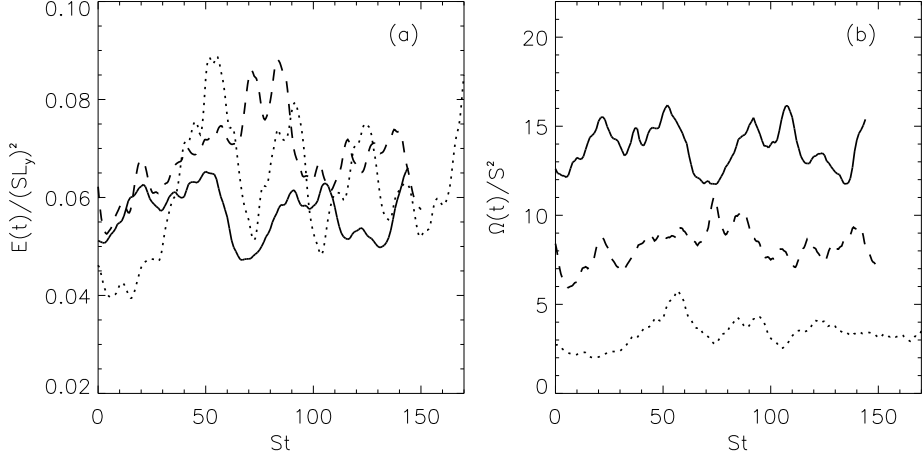}}}}
\vspace{0.5cm}
\caption{Temporal behaviour of the specific turbulent kinetic
energy $E(t)$ for three different values of Taylor Reynolds numbers,
$R_{\lambda}=59$ (dotted line), $R_{\lambda}=87$ (dashed line),
$R_{\lambda}=99$ (solid line).
(a) Specific turbulent kinetic energy $E(t)$. (b) Specific enstrophy
$\Omega(t)$.}
\label{fig2}
\end{figure}

Figure~\ref{fig2} shows the temporal evolution of the specific turbulent kinetic
energy $E(t)$ and the specific enstrophy $\Omega(t)$ defined as
%-------------------------------------------------------------------------------
\begin{equation}
E(t)=\frac{1}{2}\langle (u'_i)^2\rangle_V\,,\;\;\;\;\;
\Omega(t)=\frac{1}{2}\langle (\omega_i)^2\rangle_V\,,
\end{equation}
%-------------------------------------------------------------------------------
where $\langle\cdot\rangle_V$ denotes an average over the volume.  The vorticity
is defined as ${\bf\omega}={\bf\nabla\times u}'$.  The graphs indicate that a
statistically stationary state is established in all cases.  Additionally we
analyzed the ratio of the standard deviations $\sigma(E)$ and $\sigma(\Omega)$
to their corresponding temporal means, $\langle E\rangle_t$ and
$\langle\Omega\rangle_t$.  Both ratios were found to be smaller in comparison to
the results of remeshing simulations.  Pumir (1996) as well as Gualtieri {\it et
al.}  (2000) reported ratios of about 50$\%$ for both, energy and enstrophy at
$R_{\lambda}\sim 90$.  In contrast, we found a decrease of the ratios from about
$23\%$ at $R_{\lambda}=59$ to $10\%$ at $R_{\lambda}=99$ (see table 2).  The
large fluctuations may thus be caused by the periodic driving due to the
remeshing, though this question still remains open.
%---------------------------------------------------------------------------
%---------------------------------------------------------------------------
Statistical stationarity was further checked by comparing the ratio of turbulent
production, $P=-\langle u'_x u'_y\rangle \partial\langle u_x\rangle/\partial y$,
to the energy dissipation rate, $\epsilon=\nu\langle\left(\partial u'_i/\partial
x_j\right)^2\rangle$.  The ratio $P/\epsilon$ was always found to be very close
to 1 (see table~2).  The shear parameter $S^{\ast}=S q^2/\epsilon$ with
$q^2=\langle(u'_i)^2\rangle$ remained nearly unchanged in the five simulations,
consistent with a nearly constant ratio $-\langle u'_x u'_y\rangle/q^2$, which
should equal $1/S^{\ast}$ in the stationary case.

Next we investigated the energy dissipation rate as a function of the Taylor
Reynolds number.  The expectation is that the energy dissipation rate at high
Reynolds numbers is independent of viscosity, and is a constant of order unity
when rescaled on the integral scale $L$ and the root-mean-square velocity
$u'_{x,rms}$, i.e. $\tilde{\epsilon}=\epsilon\,L/(u'_{x,rms})^3$.
In table~2 the results are given for the present data.  Quantitiy
$\tilde{\epsilon}$ seems to saturate at order unity for $R_{\lambda}$ between 80
and 100 which cannot be exceeded with the present grid resolution.  A similar
behaviour was found in the case of homogeneous isotropic turbulence (Sreenivasan
1998).

As listed also in table~2, about half of the total amount of velocity
fluctuations is contained in the downstream component $u'_x$.  The downstream
fluctuations $\langle(u'_x)^2\rangle(t)$ were found to have always a similar
temporal shape as the total kinetic energies while the mean square moments for
the both remaining components $u'_y$ and $u'_z$ have smaller variations with
respect to time.  We observed also that the fraction
$\langle(u'_x)^2\rangle/q^2$ with $q^2=\langle(u'_i)^2\rangle$ decreased slowly
with respect to $R_{\lambda}$.  Both aspects seem to be connected to the
stronger fragmentation of coherent streaks and vortices (Schumacher \& Eckhardt
2000, 2001) that is observed with increasing $Re$ and the reduced downstream
correlation of the velocity fluctuations.
%---------------------------------------------------------------------------
\begin{table}
\begin{center}
\begin{tabular}{lccccc}
Run No. & 1 & 2 & 3 & 4 & 5 \\
\hline
$\langle (u'_x)^{2}\rangle/q^2$ & 0.57& 0.52& 0.55& 0.51& 0.51\\
$\langle (u'_y)^{2}\rangle/q^2$ & 0.12& 0.16& 0.15& 0.18& 0.18\\
$\langle (u'_z)^{2}\rangle/q^2$ & 0.31& 0.32& 0.30& 0.31& 0.31\\
$-\langle u'_x u'_y\rangle/q^2$ & 0.11& 0.12& 0.11& 0.12& 0.12\\
$S^{\ast}$ & 7.32& 8.18& 8.58& 8.22& 8.37\\
$S\tau_{\eta}$ & 0.38& 0.25& 0.25& 0.19& 0.19\\
$P/\epsilon$ & 1.04& 1.00& 0.99& 1.01& 1.01\\
$\tilde{\epsilon}$ & 1.18& 0.87& 0.70& 0.77& 0.72\\
$\sigma(\tilde{E})/\langle\tilde{E}\rangle_t$&0.20&0.12&0.11&0.09&0.10\\
$\sigma(\tilde{\Omega})/\langle\tilde{\Omega}\rangle_t$&0.23&0.13&0.12&0.09&
0.10\\
\end{tabular}
\label{tab2}
\caption{Physical properties of the homogeneous shear flow runs. Compared are
the Reynolds stress components, $\langle u'_i u'_j\rangle$, shear 
parameter, $S^{\ast}$, the ratio of Kolomogorov time $\tau_{\eta}$ to shear
time
$S^{-1}$, the ratio of turbulence production to dissipation, $P/\epsilon$ and
the dimensionless energy dissipation rate $\tilde{\epsilon}$.
The last two rows
list the ratio of standard deviations to the temporal means for the specific
turbulent kinetic energy, $\tilde{E}=E/(S L_y)^2$, and the enstrophy,
$\tilde{\Omega}=\Omega/S^2$.}
\end{center}
\end{table}
%---------------------------------------------------------------------------
\begin{figure}
\vspace{0.5cm}
\epsfxsize=0.9\hsize
{\hskip 0.0cm{\centerline{\epsfbox{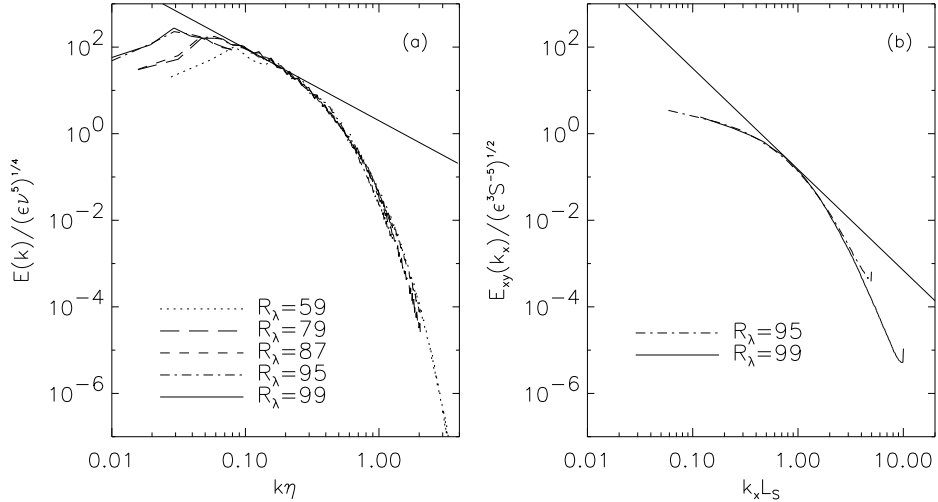}}}}
\vspace{0.5cm}
\caption{Normalized energy spectra and shear-stress cospectra for different
values of $R_{\lambda}$. (a) Energy spectra $E(k)/(\epsilon\nu^5)^{1/4}$
are plotted versus $k\eta$. 
Straight solid line indicates scaling with $k^{-5/3}$ and $C_K=2$.
(b) Shear-stress cospectra $-E_{xy}(k_x)/(\epsilon^3 S^{-5})^{1/2}$ are shown
versus $k_x L_S$ with $L_S=(\epsilon S^{-3})^{1/2}$.
The straight solid line indicates scaling with $k_x^{-7/3}$ and
$C_{xy}=0.15$. Argument $k_x$ was taken here in order to compare recent data 
to experimental findings.}
\label{fig3}
\end{figure}
%-----------------------------------------------------------------------------

In figure~\ref{fig3}a we plot the kinetic energy spectra for five different
Taylor Reynolds numbers.  With increasing values of $R_{\lambda}$ a Kolmogorov
scaling range can be observed for intermediate wavenumbers.  The straight line
is fitted with a (Kolmogorov) constant $C_K=2$. It exceeds slightly the 
values of $C_K=1.5$ -- 1.6 known from 
high-Reynolds number measurements (Pope 2000) which can be related to the 
bottleneck phenomenon (Falkovich 1994). This effect causes an 
energy pileup 
that can result in a larger $C_K$ for the small scaling ranges observed here. 
In figure~\ref{fig3}b the
cospectra are compared.  It is observed that their magnitude is always smaller
by about two orders of magnitude than the kinetic energy spectrum over the whole
range of wavenumbers, except the smallest wavenumbers.  Both shear-stress
cospectra fit to the power law proposed by Lumley (1967) in a small range of
wavenumbers.  The constant $C_{xy}$ was found here to be $C_{xy}\approx 0.15$,
equal to the values found in high-Reynolds-number experiments (Saddoughi \&
Veeravalli 1994).  As already mentioned, recent high-Reynolds number
measurements \cite{Kurien00} indicated deviations from the $k_x^{-7/3}$ decay
law.  We note here that our Taylor Reynolds number is too small to draw any
robust conclusion about the strength of deviations from the classical
Kolmogorov--Lumley scaling, but the results do not seem to contradict the 
concept
of local isotropy on the level of second order moments.

%---------------------------------------------------------------------------
\begin{table}
\begin{center}
\begin{tabular}{lccccc}
Run No. & 1 & 2 & 3 & 4 & 5 \\
\hline
$S_3(\omega_z)$ &$-7.9\times 10^{-1}$&$-7.1\times
10^{-1}$&$-6.7\times 10^{-1}$&$-6.9\times 10^{-1}$&$-6.3\times 10^{-1}$\\
$K_4(\omega_z)$ &$ 5.6\times 10^{0}$ &$ 6.4\times 10^{0} $&$
6.6\times 10^{0} $&$ 6.6\times 10^{0} $&$ 7.3\times 10^{0} $\\
$S_5(\omega_z)$ &$-8.9\times 10^{0}$ &$-1.7\times 10^{1}
$&$-1.7\times 10^{1} $&$-1.8\times 10^{1} $&$-1.9\times 10^{1} $\\
$K_6(\omega_z)$ &$ 4.9\times 10^{1}$ &$ 1.3\times 10^{2} $&$
1.5\times 10^{2} $&$ 1.5\times 10^{2} $&$ 1.9\times 10^{2} $\\
\vspace{0.3cm}
$S_7(\omega_z)$ &$-2.2\times 10^{2}$ &$-7.4\times 10^{2}
$&$-8.6\times 10^{2} $&$-8.8\times 10^{2} $&$-1.1\times 10^{3} $\\
$S_3(\partial u'_x/\partial y)$&$ 9.6\times 10^{-1}$&$ 9.2\times 10^{-1}$&$
9.0\times 10^{-1}$&$ 9.2\times 10^{-1}$&$ 8.7\times 10^{-1}$\\
$K_4(\partial u'_x/\partial y)$&$ 5.6\times 10^{0}$ &$ 6.4\times 10^{0}$ &$
6.4\times 10^{0} $&$ 7.2\times 10^{0} $&$ 7.1\times 10^{0} $\\
$S_5(\partial u'_x/\partial y)$&$ 1.0\times 10^{1}$ &$ 2.0\times 10^{1}$ &$
2.0\times 10^{1} $&$ 2.5\times 10^{1} $&$ 2.1\times 10^{1} $\\
$K_6(\partial u'_x/\partial y)$&$ 5.0\times 10^{1}$ &$ 1.3\times 10^{2}$ &$
1.3\times 10^{2} $&$ 2.0\times 10^{2} $&$ 1.7\times 10^{2} $\\
\vspace{0.3cm}
$S_7(\partial u'_x/\partial y)$&$ 2.1\times 10^{2}$ &$ 7.4\times 10^{2}$ &$
7.5\times 10^{2} $&$ 1.4\times 10^{3} $&$ 9.5\times 10^{2} $\\
$S_3(\partial u'_x/\partial x)$&$-7.4\times 10^{-1}$&$-4.4\times
10^{-1}$&$-5.0\times 10^{-1}$&$-2.7\times 10^{-1}$&$-4.0\times 10^{-1}$\\
$K_4(\partial u'_x/\partial x)$&$ 6.3\times 10^{0}$ &$ 4.8\times 10^{0} $&$
5.3\times 10^{0}$ &$ 4.3\times 10^{0} $&$ 5.0\times 10^{0} $\\
$S_5(\partial u'_x/\partial x)$&$-1.6\times 10^{1}$ &$-6.9\times 10^{0}
$&$-9.2\times 10^{0}$ &$-3.7\times 10^{0} $&$-6.6\times 10^{0} $\\
$K_6(\partial u'_x/\partial x)$&$ 1.2\times 10^{2}$ &$ 5.8\times 10^{1} $&$
7.8\times 10^{1}$ &$ 4.2\times 10^{1} $&$ 6.8\times 10^{1} $\\
$S_7(\partial u'_x/\partial x)$&$-6.1\times 10^{2}$ &$-1.5\times 10^{2}
$&$-2.6\times 10^{2}$ &$-6.9\times 10^{1} $&$-1.7\times 10^{2} $\\
\end{tabular}
\label{tab3}
\caption{Derivative moments of orders $n=3, 4, 5, 6$ and 7
of the transverse derivative
$\partial u'_x/\partial y$, the longitudinal derivative
$\partial u'_x/\partial x$
and the spanwise vorticity $\omega_z$.}
\end{center}
\end{table}
%---------------------------------------------------------------------------
\begin{figure}
\vspace{0.5cm}
\epsfxsize=1.0\hsize
{\hskip 0.0cm{\centerline{\epsfbox{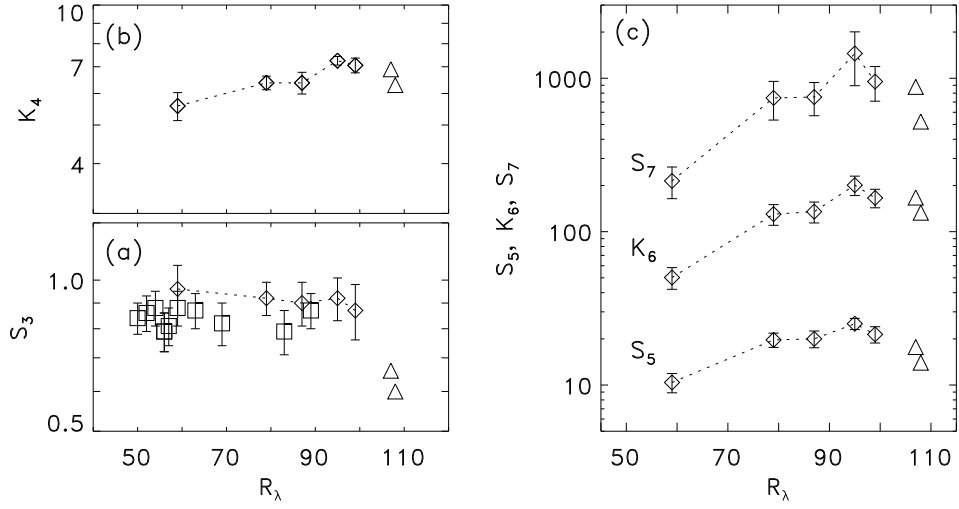}}}}
\vspace{0.5cm}
\caption{Higher order moments of the transverse derivative
$\partial u'_x/\partial y$ are compared with
other numerical and experimental findings. (a) Skewness
$S_3$. (b) Flatness
$K_4$. (c) Hyperskewness
$S_5$, hyperflatness
$K_6$ and the 7th moment
$S_7$ are compared.
$\square$, Pumir (1996); $\triangle$, Shen \& Warhaft (2000); $\lozenge$
present data. Error bars for $\lozenge$ are calculated with respect to
the variation of the data points in the $y$ direction.}
\label{fig4}
\end{figure}
%---------------------------------------------------------------------------
\begin{figure}
\vspace{0.5cm}
\epsfxsize=0.9\hsize
{\hskip 0.0cm{\centerline{\epsfbox{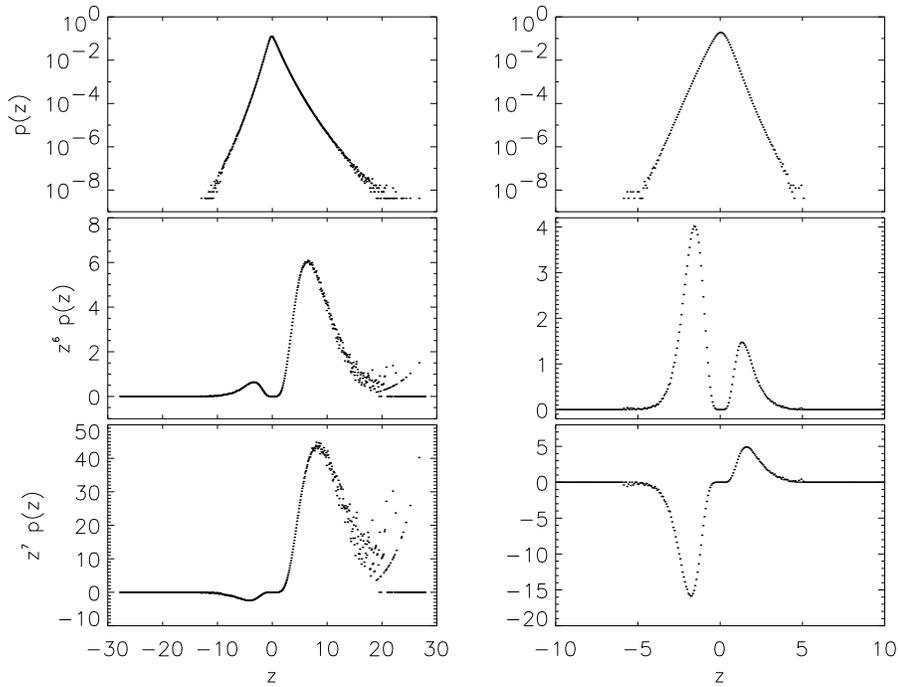}}}}
\vspace{0.5cm}
\caption{Statistical convergence of higher order derivative moments.
Left column: Results for the normalized transverse derivative
are shown, probability density function $p(z)$,
$z^6 p(z)$ and $z^7 p(z)$.
Right column: The same functions for the
normalized longitudinal derivative are plotted.
Both data sets are for $R_{\lambda}=95$ (run 4).}
\label{fig5}
\end{figure}
\section{Higher order derivative moments}
In this section, the results for higher order moments are presented.  The
quantities studied are those derivative moments that display sensitivity to
deviations from local isotropy.  Three different quantites were investigated:
the transverse derivative of the turbulent streamwise velocity, $\partial
u'_x/\partial y$, the spanwise vorticity, $\omega_z=\partial u'_y/\partial x
-\partial u'_x/\partial y$, and the longitudinal derivative of the turbulent
streamwise velocity, $\partial u'_x/\partial x$.  In table 3 we summarize the
results for orders 3 to 7 for five different simulations.  The moments of the
longitudinal derivative $\partial u'_x/\partial x$ do not vary in order of
magnitude for all values of $n$, and are found to agree with Ferchichi \&
Tavoularis (2000) as well as with Shen \& Warhaft (2000).  For a larger range of
Taylor Reynolds numbers both experiments find a slow increase for all
longitudinal moments.  If the turbulent flow would be isotropic, the odd moments
$S_{n}( \partial u'_x/\partial y)$ would be exactly zero for all odd $n\ge 0$.
In contrast, we observe nonzero values for the transverse derivative and
spanwise vorticity moments.  While the third order slowly decreases in the
range of Taylor Reynolds numbers, fifth and seventh order increase with
$R_{\lambda}$.  In figure~\ref{fig4} we compare our results with the
numerical data of Pumir (1996) for $S_{3}(\partial u'_x/\partial y)$ and with
the experimental data of Shen \& Warhaft (2000) at $R_{\lambda}\sim 100$.  The
error bars for our data display the standard deviation of the $y$ profiles of
the moments.  Averaging was always performed in time and in planes at fixed $y$.
Data points closest to the boundaries are excluded from the evaluation of the
averages.  The snapshots of the full velocity field were always separated by at
least one shear time unit, $St=1$.  The minimal number of used snapshots was 114
(for run 5).  The results are in agreement with previous numerical and
experimental findings.

The fluctuations around the mean grow for increasing order as expected.  These
fluctuations were subject to a more detailed investigation.  Figure~\ref{fig5}
shows the probability density functions (pdf), denoted as $p(z)$, of
$z=(\partial u'_x/\partial y)/\langle(\partial u'_x/\partial y)^2\rangle^{1/2}$
in the left column and of $z=(\partial u'_x/\partial x)/\langle(\partial
u'_x/\partial x)^2\rangle^{1/2}$ in the right column.  The results are
calculated from $6\times 10^8$ data points at $R_{\lambda}=95$. A check of the
statistical convergence of the $n$th order moments can be made by plotting the
integrands $z^n p(z)$ over $z$ which is also shown in figure~\ref{fig5}.  While
the longitudinal derivative has converged well, the
integrands for the transverse derivative show a strong scatter in the tails of
the pdf.  The area that covers the scattered data measures the uncertainty
around the mean in an alternative way.

In the following, we wish to discuss the strong scattering of higher order
moments and will demonstrate by means of a simple analytical model that this is
a more general feature for data analysis.  The pdf (see figure~\ref{fig5}) 
can be roughly approximated by the following normalized exponential
distribution $p(z)=(\alpha\cosh(\alpha z_0)/2)\,\exp(-\alpha |z-z_0|)$ 
for which all
moments can be evaluated analytically, following to $\langle z^n\rangle\sim
n!\,$.  Let $N$ be the number of statistically independent samples.  The
standard deviation, which is defined as $\sigma(z^n)=[(\langle z^{2n}\rangle-
\langle z^{n}\rangle^2)/(N-1)]^{1/2}$, follows then as
%----------------------------------------------------------------------------
\begin{equation}
\sigma(z^n)\simeq\pm\frac{1}{\alpha^{n}}
\left[
\frac{(2n)!-(\tanh(\alpha z_0) n!)^2}{N}\right]^{1/2}\,,
\label{sigma}
\end{equation}
%----------------------------------------------------------------------------
for the odd normalized moments $S_n=\int\,z^n p(z) dz$.  The denominator was
simplified because $N\gg 1$.  When taking, e.g., $\alpha=1$ and $z_0=1$, the
standard deviation for the seventh moment decreases from 24$\%$ at $N=10^5$ to
1$\%$ at $N=10^8$.  Thus, the calculation demonstrates clearly the sensitivity
of higher order moments on the number $N$ which can give rise to very large
fluctuations around the mean as observed in our simulations.  While moments
itself are proportional to the factorial of the order, $n!$, its
fluctuations become even proportional to $((2n)!)^{1/2}$. 
Tennekes \& Lumley (1972) estimated $N(1)=T/(2\tau)$ for a first order moment, 
where $\tau$ is the integral time scale and $T$ the duration of the measurement.  
Starting from this point, Sreenivasan {\it et al.}
(1978) extracted relations for the accuracy of higher order moments using their
atmospheric boundary layer data. Following their lines, the number of
statistically independent samples for an $n$th moment, $N(n)$, would result
in $N(n)/N(1)=(0.82-0.07n)^{-1}$.  Aside from the fact that the
fluctuations for fixed $N$ and growing $n$ can increase rapidly, this dependence
would improve the statistical convergence weakly.

In summary, we have presented an alternative method of modelling a 
statistically stationary
homogeneous shear flow by using stress-free boundary conditions in $y$ 
direction.  On the level of second order moments our results are
not in conflict with the classical concepts of local isotropy as indicated by
the power spectra of the velocity correlations.  We found
non-decreasing derivative moments for the orders 4 to 7 which agree with the
experimental findings of Shen \& Warhaft for $R_{\lambda}\sim 100$ but are
in apparent violation of local isotropy.  A next step
would be to relate typical structures of turbulent shear flows, such as
streamwise vortices and streamwise streaks, to the statistical properties
presented here.
\begin{acknowledgments}
The author thanks B.~Eckhardt, A.~Pumir, K.~R.~Sreenivasan, P.~K.~Yeung and
Z.~Warhaft for fruitful discussions and helpful comments. The numerical
simulations were done on a Cray T-90 at the John von Neumann-Institut f\"ur
Computing at the Forschungszentrum J\"ulich. This work was sponsored in parts by the Feodor-Lynen Fellowship
Program of the Alexander von Humboldt--Foundation and Yale University. 
\end{acknowledgments}

\end{document}